%% file: Stewart2015.tex
\documentclass[iop]{emulateapj}
\usepackage{graphicx}
\usepackage{amsmath}
\usepackage{amstext}
\usepackage{color}
\usepackage[backref,breaklinks,colorlinks,citecolor=blue]{hyperref} 
\usepackage[all]{hypcap} 

\usepackage{multirow}
\usepackage{longtable}
\usepackage{adjustbox}

\newcommand{\ditto}{''}
\newcommand{\tild}{\raise.27ex\hbox{$\scriptstyle\sim$}}
\newcommand{\Cassini}{\textit{Cassini~}}
\newcommand{\CassiniVIMS}{\textit{Cassini}-VIMS~}

\shorttitle{\Cassini Atlas Of Stellar Spectra}
\shortauthors{Stewart et. al.}

\begin{document}

\title{An atlas of bright star spectra in the near infrared from \CassiniVIMS}
\author{Paul N. Stewart and Peter G. Tuthill}
\affil{Sydney Institute for Astronomy}
\affil{School of Physics, The University of Sydney, NSW 2006, Australia}
\email{p.stewart@physics.usyd.edu.au}
\author{Philip D. Nicholson}
\affil{Department of Astronomy}
\affil{Cornell University, Ithaca, NY 14853, USA}
\author{G. C. Sloan}
\affil{Cornell Center for Astrophyics and Planetary Science}
\affil{Cornell University, Ithaca, NY 14853, USA}
\affil{Carl Sagan Institute}
\affil{Cornell University, Ithaca, NY 14853, USA}
\and
\author{Matthew M. Hedman}
\affil{Department of Physics}
\affil{University of Idaho, Moscow, ID 83844, USA}

\label{firstpage}

\begin{abstract}
We present the \Cassini Atlas Of Stellar Spectra (CAOSS), comprised of near-infrared low-resolution spectra of bright stars recovered from space-based observations by the \Cassini spacecraft.
The 65 stellar targets in the atlas are predominately M, K and S giants.
However it also contains spectra of other bright nearby stars including carbon stars and main sequence stars from A to F.
The spectra presented are free of all spectral contamination caused by the Earth's atmosphere, including the detrimental telluric molecular bands which put parts of the near-infrared spectrum out of reach of terrestrial observations.
With a single instrument, a spectro-photometric dataset is recovered that spans the near-infrared from 0.8 to 5.1 microns with spectral resolution ranging from R=53.5 to R=325.
Spectra have been calibrated into absolute flux units after careful characterisation of the instrumental spectral efficiency.
Spectral energy distributions for most stars match closely with literature values.
All final data products have been made available online.

\end{abstract}

\section{Introduction}


The recovery of high quality stellar spectra underpins an integral part of our understanding of stellar composition, evolution, and temporal behaviour.
Unfortunately important molecular spectral features expected to be present are masked by the existence of the same molecular species in our own atmosphere.
This is especially true for cooler evolved stars, where substantial quantities of H$_2$O and CO$_2$ are known to form.
This results in large parts of the infrared region being effectively opaque to ground-based observations.
Piercing this veil is only possible by observing beyond the Earth's atmosphere from space-based platforms.

There has yet to be a space-based spectrometer dedicated to surveying bright stars over a broad range of infrared wavelengths, despite the potential scientific merit and the severe obstacles to observing from the ground.
However, the Visual and Infrared Mapping Spectrometer (VIMS) instrument on the \Cassini spacecraft has measured the spectra of many bright stars in the wavelength range of 0.8-5.1 microns.
This broad wavelength range is distinct from any other existing space-based observatory platform.
The International Ultraviolet Explorer (IUE) provides a survey of spectra shorter than this range, spanning 0.11 to 0.32\,$\mu m$~\citep{Boggess1978}.
The longer wavelengths overlap with surveys by the Short Wave Spectrometer (SWS) on ESA's Infrared Space Observatory (ISO) covering 2.38 to 45.2\,$\mu m$~\citep{DeGraauw1996}.
The Space Telescope Imaging Spectrograph (STIS) on the Hubble Space Telescope (HST) can observe spectra between 0.11 and 1.0\,$\mu m$~\citep{Woodgate1998}, while the Near Infrared Camera and Multi-Object Spectrometer (NICMOS) on HST was capable of observing the remaining 1.0 - 2.5\,$\mu m$ range at the shorter end of the infrared ~\citep{Thompson1998}.
However, these HST instruments have not been used for survey purposes and therefore only specifically targeted objects have been observed.
Photometry in the \textit{H}, \textit{K}, \textit{L}, and \textit{M} bands was acquired for 92 of the brightest stars by the Diffuse Infrared Background Experiment (DIRBE) on COBE~\citep{DIRBE1998}.
A much more extensive catalogue of ground based photometric observations was produced from 2MASS observations in the \textit{J}, \textit{H}, and \textit{Ks} bands~\citep{Skrutskie2006}.
Observations at longer wavelengths have been performed by SPITZER ($>$ 5.3\,$\mu m$), MSX ($>$ 8.3\,$\mu m$), IRAS ($>$ 12\,$\mu m$), AKARI ($>$ 50\,$\mu m$), and Herschel ($>$ 55\,$\mu m$).
None of these instruments span the entire range accessible to VIMS and many of them have been unable to observe the brightest stars as they are beyond the instrumental saturation limit.
VIMS has previously been used for stellar studies using stellar occultations by Saturn's rings~\citep{Stewart2013, Stewart2014a, Stewart2015_cwleo, Stewart2015_mira, Stewart2015_tomo}, but until now has not been used for studies of stellar spectra.

The atlas provides unique access to broadband, space-based spectra of many of the brightest stars in the sky, albeit at a relatively low spectral resolution.
This paper starts with a description of the experiment and instrument in Section~\ref{sec:expdesc}, followed by details on the data reduction processes in Section~\ref{sec:datared}.
Sample spectra are presented in Section~\ref{sec:samples}, and compared with those obtained by ISO and IRTF, and photometry from DIRBE and 2MASS.
The full list of stars in the atlas is provided in Appendix~\ref{app:caoss} and reduced spectra are available online at \url{http://www.physics.usyd.edu.au/sifa/caoss/} and Vizier.
By making these spectra available, we anticipate scientific benefits beyond the goals of the \Cassini mission, and hope to enable new sections of the astronomical community to gain directly from a planetary exploration mission.

\section{Experiment Description}\label{sec:expdesc}

The observations were originally acquired for the purposes of monitoring for any temporal variations within the VIMS instrument and to provide a baseline for the interpretation of stellar occultation results.
Their acquisition started in the early stages of the mission, whilst the spacecraft was in ``cruise'' and on its way to Saturn, with the program expected to continue through to the end of the mission.
The observations from this monitoring program were not intended to be used for stellar science.
However, they fill a void in spectral coverage for a sufficient number of bright stars so as to warrant dissemination to the wider astronomical community.

\subsection{The VIMS Instrument}
VIMS was designed to observe the spatially resolved visible and near-infrared spectra of various surfaces in the Saturnian system.
An overview of the instrumental design was published by \citet{Brown2004} and the significant points will be reviewed here.
The instrument has two independent, bore-sighted telescopes feeding separate visible and near-infrared spectrometers referred to as VIMS-VIS and VIMS-IR respectively.
The diameters of these telescopes are 4.5\,cm for the visible and 23\,cm for the infrared, with the corresponding spectrometers spanning 0.35 - 1.05\,$\mu m$ in 96 channels and 0.8-5.1\,$\mu m$ in 256 channels.
Both channels were operating when the observations used in this atlas were acquired, but due to the small collecting area  of its telescope the visible channel generally did not yield sufficient signal to noise ratio to provide useful stellar spectra.
Consequently, the information from the visible channel was not used in the preparation of this atlas, and only data from VIMS-IR is presented.
VIMS-IR uses a triply-blazed diffraction grating with a spectrometer of conventional design with the spectral resolution (R) changing linearly from 53.5 at the blue end to 325 at the longest wavelength.
The detector response was measured in ground tests prior to launch and was found to be entirely linear with incident flux.
Due to its large spectral range, the infrared spectrometer employs a 4-element blocking filter to exclude higher-order transmissions from the diffraction grating; this results in small gaps in the spectra near 1.6, 3.0, and 3.9\,$\mu m$ due to the boundaries between the filter segments.
The filter gap at 3.9\,$\mu m$ affects 4 spectral channels, whilst the other two gaps are each two spectral channels wide.
Because of the 20-year duration of the \Cassini mission, cryogens are impractical and VIMS is passively cooled via a radiator.
The 256-element InSb alloy detector operates at a stable temperature of 58-60\,K, but the spectrometer optics are typically at \tild130\,K.
The instrumental thermal background is measured periodically during routine observations and subtracted from the target spectrum.

The VIMS instrument was designed to observe extended objects that entirely fill its instantaneous field of view, such as Saturn, its moons, and the ring system.
Images are built up by measuring the spectrum of a single pointing individually and then moving the tip and tilt of the secondary mirror to raster in two dimensions over the required field.
Each ``pixel'' in this time-series image scan subtends a relatively large 228$\times$493\,$\mu$rad rectangular piece of the sky, with the tip and tilt of the secondary mirror providing a measured pixel centre separation of 250$\times$500\,$\mu$rad~\citep{Brown2004}.
This results in a photon collection area covering only \tild90\% of the image.
The spacecraft pointing accuracy has been determined to be $<$629\,$\mu$rad and the pointing stability has been measured to plateau below 6\,$\mu$rad~\citep{Pilinski2009}.
This is not a concern for existing VIMS observations within the Saturnian system which involve filled pixels, but it introduces problems when observing unresolved point-sources such as stars.
As the pixel size is much larger than the stellar PSF, it is likely that \tild10\% of the time the stellar image will fall partially within these uncollected inter-pixel dead gaps in the 2D raster scan.
In spite of this uncertainty in the precise position of the star within a pixel, drift in pointing over the duration of an observation is negligible.
There are known to be small variations in the size and shape of this instrumental pixel as a function of wavelength.
Observations in which the PSF falls substantially within these unmeasured inter-pixel dead zones are identified by fitting a modelled instrumental PSF to the observation (P. Stewart 2015 PhD Thesis in preparation).
Such observations were found to be spectrally and photometrically inconsistent, and have been identified and removed from the main atlas.

\citet{McCord2004} 
show that under-filled pixels can effect the wavelength calibration by up to half a spectral channel (6-8\,nm) and more recently there has been a report of a gradual monotonic drift in wavelength of up to 9\,nm toward the red since the spacecraft's launch~\citep{VIMS2015}.
This has been incorporated into the wavelength uncertainty discussed in Section~\ref{sec:data_products}.

VIMS-IR images used in this project have been acquired in one of two imaging modes, known as NORMAL and HIRES.
The NORMAL mode co-adds two adjacent rectangular pixels to produce square pixels, whilst HIRES keeps the instrumental rectangular pixels to produce higher resolution images in one dimension.
The two modes have been shown to provide consistent spectra, and data have been averaged together where observations in both modes exist.


\section{Data Reduction Process}\label{sec:datared}

The observations were obtained from the data archive at the Planetary Rings Node of NASA's Planetary Data System (PDS)~\citep{NASA2015}.
The raw format stores the intensity across the spectrum in background-subtracted detector counts together with the background itself.
The parameters of the observation and instrumental operation are recorded in the data file headers.

\begin{figure}[t]
 \centering
 {\includegraphics[width=\columnwidth,clip=true,trim=0 30 0 50]{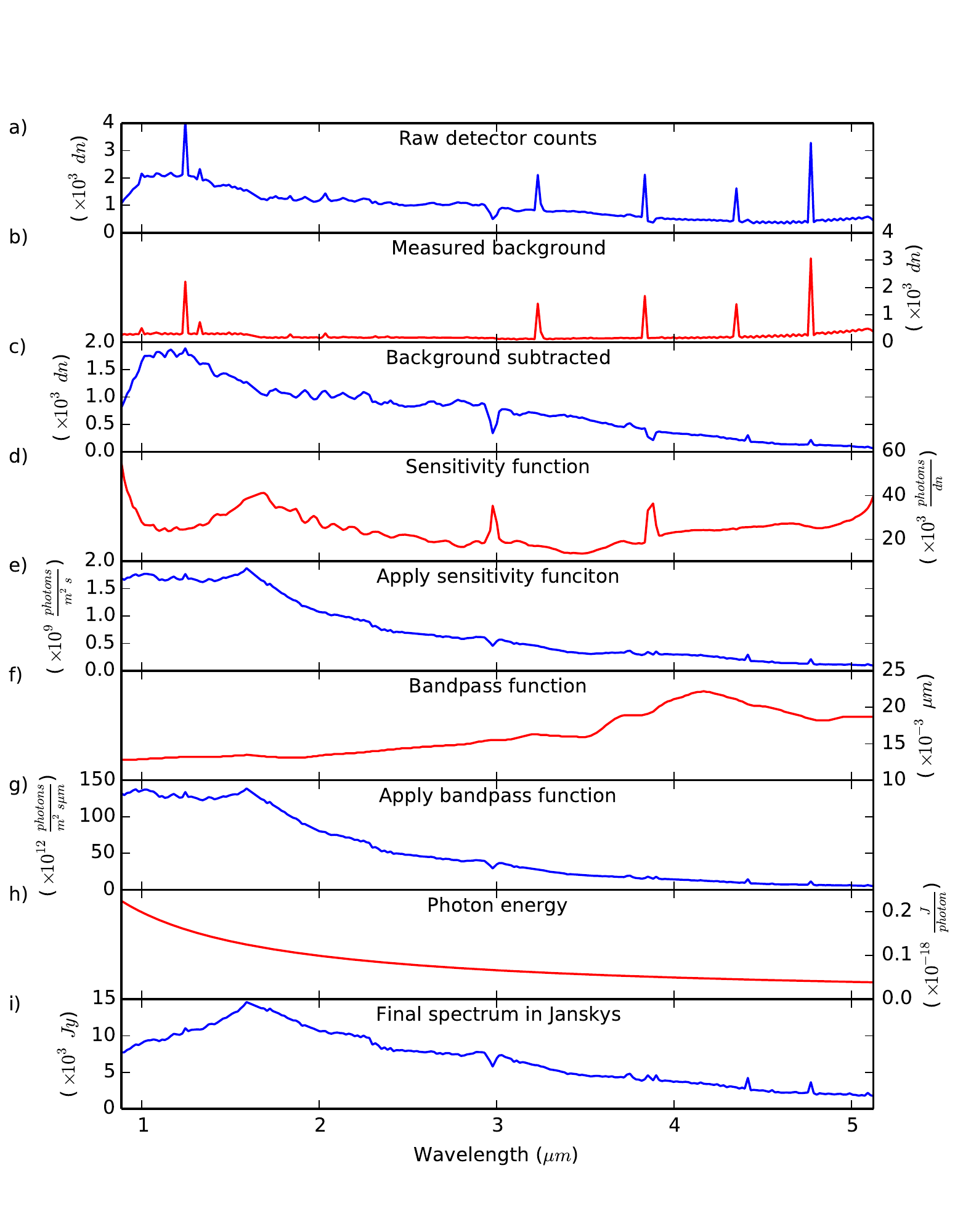}}
 \caption{The data reduction process illustrated for a single 320 ms exposure of $\alpha$ Boo in sequence C34.
 The blue curves show the data progressing through the reduction process, whilst the requisite calibration curves are in red.
 Full details of each panel are in Section~\ref{sec:datared}.
 }
 \label{fig:data_reduction}
\end{figure}

These observations were performed in the raster-scanning mode described above and stored as 3D image ``cubes''.
The full 0.8-5.1\,$\mu m$ spectrum was obtained simultaneously for each signal pixel individually, with typical integration times of 320 or 640 ms per pixel.
The images recorded by VIMS usually cover a 8x8 or 12x12 pixel field in all 352 spectral channels using measurements from both VIMS-VIS and VIMS-IR, but only the 256 channels recorded by the infrared instrument are used here.
From the recorded data (Figure~\ref{fig:data_reduction}a), VIMS subtracts the measured background (Figure~\ref{fig:data_reduction}b) before transmitting them both back to Earth.
The first step in the extraction of the stellar spectra from these cubes is to sum over a 3x3 pixel square, centred on the brightest pixel, producing a base spectral measurement as shown in Figure~\ref{fig:data_reduction}c.
In order to determine if any spectral channels saturated during the exposure, the measured and separately recorded background spectrum (Figure~\ref{fig:data_reduction}b) is added to the brightest pixel's counts (Figure~\ref{fig:data_reduction}a).
Any spectral channels which now show 4095 counts are flagged as saturated and are omitted.
These occur frequently at the hot pixels shown as spikes in Figure~\ref{fig:data_reduction}b, such as the saturated pixel at 1.25\,$\mu m$ in Figure~\ref{fig:data_reduction}a.
Similarly, pixels affected by cosmic rays are identified and flagged for omission.
These base spectra are divided by the geometric area of the telescope's aperture and the exposure time of the observation (extracted from data headers) to give the measured spectra in counts/m$^2$/s.
The instrumental sensitivity function (Figure~\ref{fig:data_reduction}d), combining both the electronic gain and the detector sensitivity as a function of wavelength, is used to convert this into photons/m$^2$/s (Figure~\ref{fig:data_reduction}e).
By dividing by the instrumental bandpass shown in Figure~\ref{fig:data_reduction}f ($\Delta\lambda$), we arrive at a spectrum in photons/m$^2$/s/$\mu m$ (Figure~\ref{fig:data_reduction}g) which is multiplied by the photon energy (Figure~\ref{fig:data_reduction}h) to give J/m$^2$/s/$\mu m$.
Finally this is converted into Janskys as shown in Figure~\ref{fig:data_reduction}i ($=10^{-26}Wm^{-2}Hz^{-1}$) in order to compare with previously published spectra from the literature.

The sensitivity and bandpass functions were derived as part of the VIMS-IR calibration for extended sources.
\citet{Clark2012} details the development of these functions through to the latest version (RC17) which is used in this work.
For stellar sources which do not fill the spectrometer's pixel, the sensitivity function is expected to differ somewhat from these standard values.
A wavelength-dependant systematic deviation between the recovered VIMS spectra and those previously published was noted in this work, and was found to be consistent across a range of stars of various types and spanning multiple epochs.
A correction ratio was constructed based on a comparison of VIMS spectra for standard stars to corresponding literature spectra from \citet{Engelke2006} and \citet{Rayner2009}.
The ratios of sections of the literature spectra to VIMS observations were averaged to produce a mean brightness ratio by which VIMS observations were multiplied.
Variable stars known to exhibit strong wavelength-dependent temporal variations which have the potential to introduce spectral artefacts, and those with an amplitude of greater than one magnitude in any part of the near infrared have been omitted.
These relevant reference spectra are summarised in Table~\ref{tab:correction}.
The final correction function is available online with the atlas, and is provided to enable the application of alternative calibrations.

\begin{figure}[t]
 \centering
 {\includegraphics[width=.99\columnwidth,clip=true,trim=0 10 0 10]{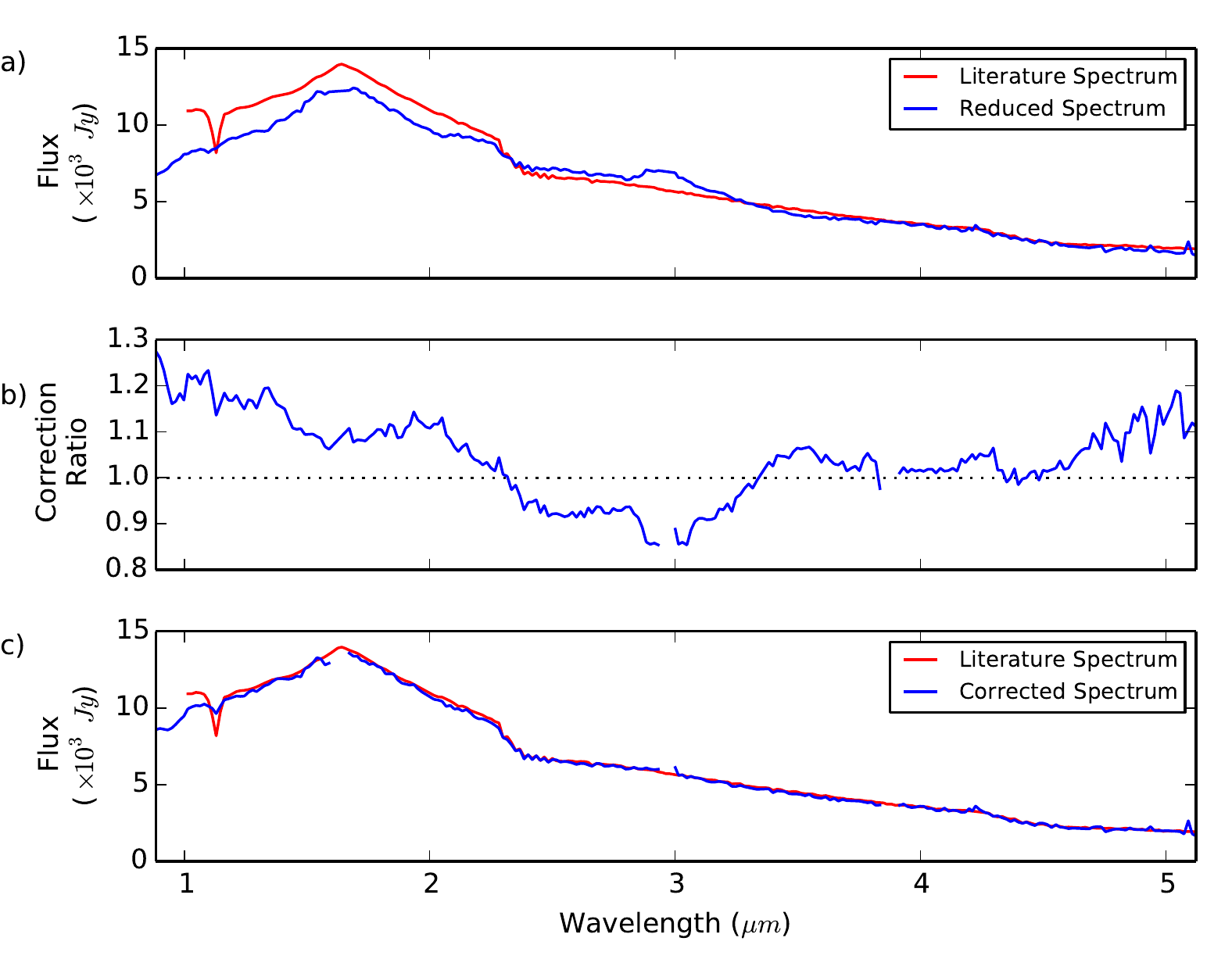}}
 \caption{A demonstration of the effectiveness of the correction function discussed in Section~\ref{sec:datared} on the S16 observation of $\alpha$ Boo.
 Panel (a) shows a comparison of the processed VIMS spectrum (in blue) against a published spectrum from \citet{Engelke2006}, resampled to the same spectral resolution.
 Panel (b) shows the correction ratio by which all VIMS stellar spectra are multiplied to produce a corrected spectrum.
 The final panel (c) shows the corrected VIMS spectrum against the resampled \citet{Engelke2006} curve.
 }
 \label{fig:correction}
\end{figure}

Figure~\ref{fig:correction} demonstrates the effectiveness of this correction as applied to $\alpha$ Boo from sequence S16.
The resultant corrected spectrum provides a good match to the general shape of the literature data, as well as some of the finer details, especially through 2.2-2.6\,$\mu m$ and the narrow absorption feature near 1.1\,$\mu m$.
For comparison purposes, the literature spectra have been downsampled to match the spectral resolution of VIMS.
In order to to ensure the robustness of this comparison we also produced a calibration curve omitting the `test' epoch and found the resultant curve to be indistinguishable from that produced using all observations which have literature spectra.
Application of this correction function is only necessary when observing point-sources, and calibrating filled-pixel spectra does not require this step.

The spectra are presented without any correction for interstellar reddening for the following reasons.
The majority of sources in the atlas are close to the Sun and thus have A$_{\text{V}}$ values of approximately zero, with half the sample within 110 pc and only 10 stars further than 500 pc.
Extinction corrections are much smaller in the infrared than in the optical, reducing the impact of interstellar reddening on stars in the atlas.
It is not envisioned that the calibration pipeline presented here will change, whereas extinction corrections, dependent on poorly known distances and estimated extinctions to that distance along a given line-of-sight, will almost certainly change in the future. This is especially likely as 3D extinction models of the Galaxy improve over the coming decades.


\begin{table}
\centering
\caption{A table of observations used to derive the calibration correction ratio.
The first set of data are derived from space-based ISO SWS observations, whilst the second set originate from terrestrial observations performed with IRTF.
Epochs are in \Cassini planning sequences, and are available in UTC and Julian dates in Table~\ref{tab:calibrated_data}.\label{tab:correction}}
\begin{tabular}{c l c}
  $\lambda$ ($\mu m$)	& Stars Used (Epochs)		& Source for Ref. Spectrum\\
\hline
  2.4 - 5.1 		& alp Boo (C27, C34, S36) 	&\citet{Engelke2006}	\\
			& alp Tau (C35) 		&\ditto			\\
			& bet UMi (S35) 		&\ditto			\\
			& alp Cet (S18) 		&\ditto			\\
			& bet And (S18) 		&\ditto			\\
			& gam Cru (C35, S38, S53) 	&\ditto			\\
			& alp CMa (C39, S69) 		&\ditto			\\

 0.8 - 2.4		& alp Boo (C27, C34, S36)	&\citet{Rayner2009} 	\\
			& alp Ori (C33, S18) 		&\ditto			\\
			& mu Cep (S22)	 		&\ditto			\\
			& R Lyr (S13, S79) 		&\ditto			\\
			& alp Her (S13, S53) 		&\ditto			\\
			& rho Per (S18) 		&\ditto			\\

\end{tabular}

\end{table}

\subsection{Data Products}\label{sec:data_products}

Spectra are identified by the name of the stellar target and the mission sequence in which they were observed.
These sequences are labelled either ``C'' or ``S'' identifying either ``cruise'' or ``Saturnian orbit''.
They are then numbered sequentially with time with each ``S'' sequence spanning several planetary orbits by the spacecraft.
As an example, RDor C33 is a spectrum of the star R Doradus observed in the 33rd cruise sequence.
The UTC and Julian dates of these are included in Table~\ref{tab:calibrated_data}, as are the number of exposures and cumulative integration time of all exposures for each observation.

Each spectrum presented is the average of all frames acquired within a given observation.
The actual UTC date and time of each of the averaged exposures is included in the data header, along with the exposure times, image dimensions, and instrument mode.
The final data products are available as simple text files, and as four column fits tables, with the columns being centre wavelength, flux density, uncertainty in centre wavelength and uncertainty in flux density.

\begin{table}[t]
\centering
\caption{The number of stars of each spectral class and variability type in CAOSS}
\begin{tabular}{| l | l r |}
\hline
By Spectral Class&	O&		1\\
&			A&		2\\
&			F&		2\\
&			C&		3\\
&			S&		3\\
&			K&		8\\
&			M0-4&		17\\
&			M5-9&		29\\
  \hline
By Variability Type&	Irregular&	13\\
&			Mira&		16\\
&			Semiregular&	20\\
\hline
\end{tabular}
\label{tab:data}
\end{table}

Table~\ref{tab:starlist} includes optical spectral classes for each source, based on a review of the classification literature.
We generally report the oldest classification(s) consistent with subsequent papers, except where improved classifications are available from \cite{M73}, \cite{K74}, or \cite{K89}.
\cite{Sloan2015} report spectral types for 14 targets in the CAOSS sample, based on slightly different rules, and as a consequence several differ by a step in spectral class or luminosity class.
These differences indicate the level of uncertainty in the classifications.
We do not report spectral types for optical companions which do not contribute significantly to the near-infrared spectrum.

\section{Sample Spectra}\label{sec:samples}

The spectra produced in this work include stars from many spectral classes including A and G main sequence stars and K giants, but due to engineering requirements for bright targets, the atlas is dominated by evolved M giants.
Many of these targets have complex molecular atmospheres and exhibit variability.
This section shows some of the recovered spectra and compares them, where possible, to existing literature measurements.
These literature values come predominately from space-based ISO SWS spectra~\citep{Engelke2006} and ground-based IRTF spectra~\citep{Rayner2009}, with space-based photometry from DIRBE~\citep{DIRBE1998}.
WISE photometry was found to be unreliable for many of these targets due to unflagged saturation.
For accurate comparison, both the ISO and IRTF spectra have been downsampled to match the spectral resolution of VIMS.

\begin{figure}[tb]
 \centering
 {\includegraphics[width=.9\columnwidth,clip=true,trim=0 28 0 50]{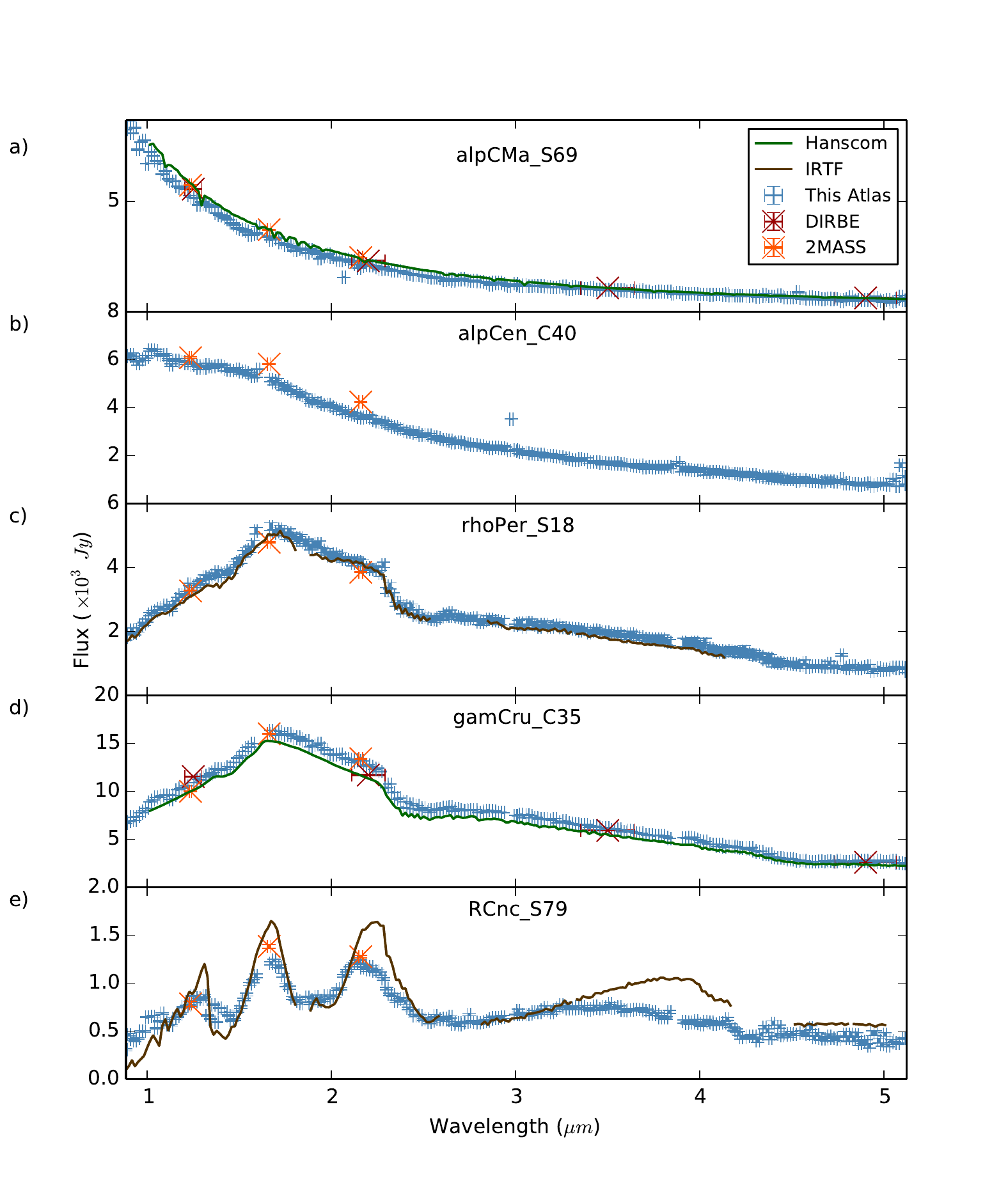}}
 \caption{A comparison of spectra presented in this atlas (blue) to the literature.
 The vertical axis is stellar flux density in $10^3$ Janskys and the horizontal axis is wavelength in microns.
 Spectra of Hansom standard stars from ISO by \citet{Engelke2006} are shown as green curves and IRTF spectra by \citet{Rayner2009} are rendered as brown curves.
 Photometric measurements are shown as individual points with error bars in dark red and orange for DIRBE and 2MASS, respectively.
 The discrepancy in panel (e) is due to the inherently variable nature of the target star, R Cnc.}
 \label{fig:sample_spectra}
\end{figure}

To give a sample of the CAOSS data, Figure~\ref{fig:sample_spectra} shows reduced spectra for five different stellar targets.
Each panel labelled (a) to (e), contains an observation at a single epoch for a single star.
Panel (a) shows the A1 star Sirius ($\alpha$ Canis Majoris) with a spectrum comprising predominately the Rayleigh-Jeans side of a hot Planck curve (Sirius has an effective surface temperature of approximately 10,000~K with peak intensity in the ultraviolet).
This shows very good agreement with the \citet{Engelke2006} spectrum, as well as both the DIRBE and 2MASS photometry.
Panel (b) contains the blended spectrum of the binary system, $\alpha$ Centauri A and B and also shows good agreement to the 2MASS photometry.
Panels (c) and (d) contain the early-mid M giants $\rho$ Persei and $\gamma$ Crucis.
These spectra have broadly similar shapes, yet very different magnitudes due to the latter's relative proximity.
They agree well with existing literature spectra and photometry from all four sources.
The final panel (e), shows R Cnc, a Mira variable type star with a spectral class of M6.5-9.
Such stars are known to exhibit huge variations in magnitude over a period on the order of a year.
The spectrum presented only partly agree with the published \citet{Rayner2009} IRTF spectra and 2MASS photometry, which in keeping with the known spectral variability of such objects.

\begin{figure}[tb]
 \centering
 {\includegraphics[width=.99\columnwidth,clip=true,trim=0 0 0 25]{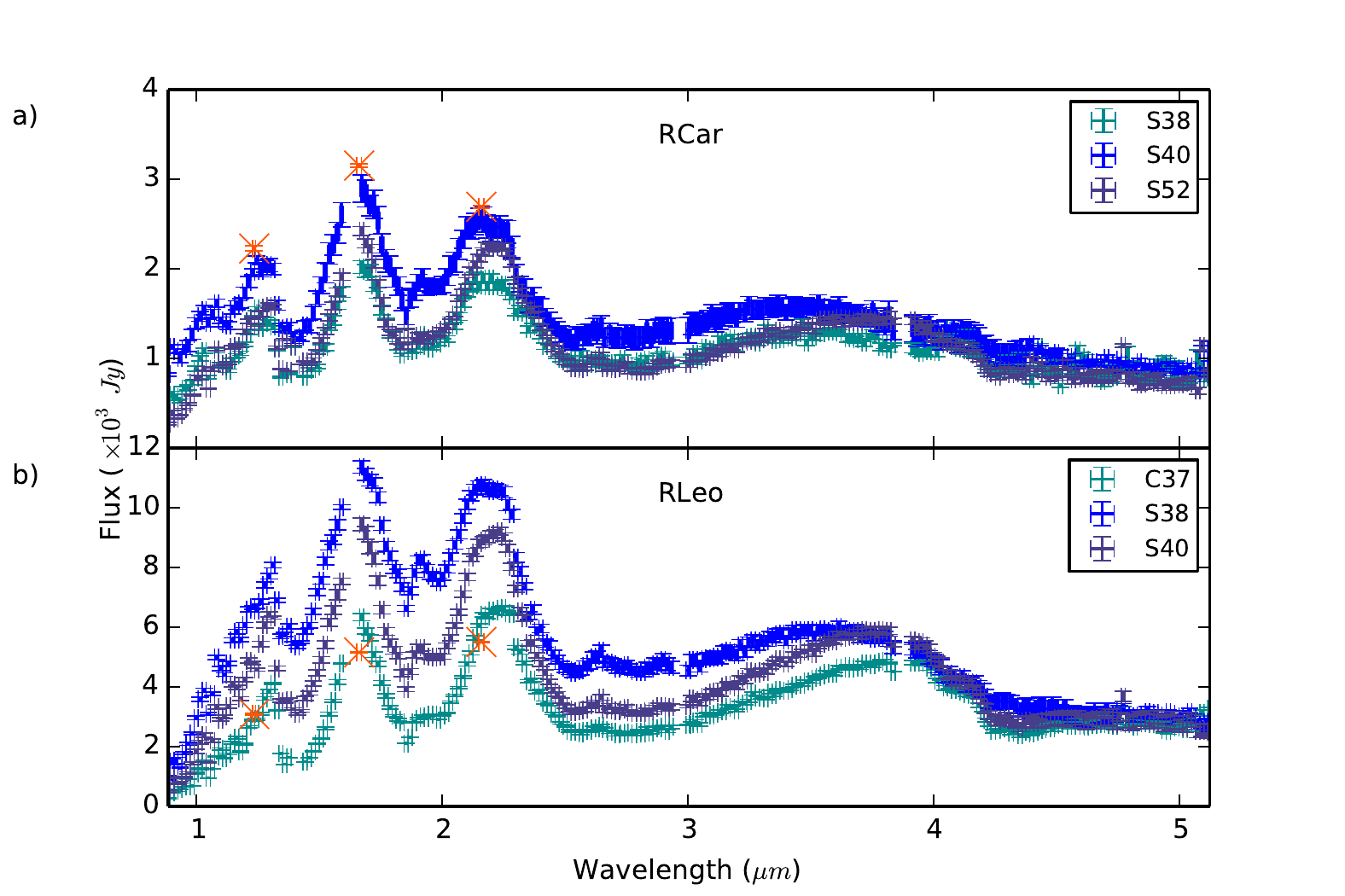}}
 \caption{Multiple epochs of variable stars showing significant temporal variation.
 Each panel shows a spectrum for the three epochs indicated in the legend.
 The orange points show 2MASS photometry from an epoch independent of the VIMS observations.}
 \label{fig:sample_spectra_var}
\end{figure}

Figure~\ref{fig:sample_spectra_var} shows two more Mira variable stars for which there are multiple epochs presented in this atlas.
Large changes in overall flux, accompanied by more minor shifts in spectral shape are noted between epochs.
The variations in flux for these particular stars, as with all variable stars in the atlas, have been checked against the AAVSO lightcurves and found to be broadly consistent, but no detailed analysis of stellar variability has been performed in this work.

\section{Conclusions}

We have produced a database of space-based near-infrared spectra for 73 stellar targets at 109 epochs.
These spectra have been made publicly available online in flexible data formats.
Where comparison was possible they have been shown to be consistent with existing spectra and photometry.
In spite of the relatively low spectral resolution, we believe that these spectra are of sufficient quality to continuously bridge a vital yet unsurveyed part of the spectrum and to become a valuable resource for the stellar astrophysical community.
As spectra from this dataset span the telluric water bands they particularly enable the study of evolved stars, known to exhibit strong spectral structure from their own atmospheric water.
Spectra presented in this atlas have already been used in a study of the atmosphere of Mira, including an assessment of models of its atmospheric behaviour~\citep{Stewart2015_tomo}.

\section*{Acknowledgements}
This research made use of the SIMBAD database, operated at CDS, Strasbourg, France.

Some of the data presented in this paper were obtained from the Mikulski Archive for Space Telescopes (MAST). STScI is operated by the Association of Universities for Research in Astronomy, Inc., under NASA contract NAS5-26555. Support for MAST for non-HST data is provided by the NASA Office of Space Science via grant NNX13AC07G and by other grants and contracts.

This publication makes use of data products from the Two Micron All Sky Survey, which is a joint project of the University of Massachusetts and the Infrared Processing and Analysis Center/California Institute of Technology, funded by the National Aeronautics and Space Administration and the National Science Foundation.

\bibliographystyle{apj} 
\bibliography{library}

\appendix
\section{The \Cassini Atlas Of Stellar Spectra}\label{app:caoss}

The details of all stars contained in CAOSS are listed in Table~\ref{tab:starlist}, including coordinates, spectral type and variability type where relevant.
A full list of flux density calibrated spectra to date is contained in Table~\ref{tab:calibrated_data}.
Observations processed identically to those in Table~\ref{tab:calibrated_data} yet containing significant deviations from the literature values are presented online with the atlas, but not listed here.
These spectra are still expected to be of value in situations where an absolute flux density calibration is not required.

\clearpage
\include{starlist}

\include{caoss}


\end{document}

%% file: starlist.tex
\begin{center}
\begin{longtable}{lllcrrlll}
\caption{Stars in the Cassini Atlas of Stellar Spectra. Details of individual observations are listed in Table~\ref{tab:calibrated_data}.\label{tab:starlist}}
\\ \multicolumn{3}{c}{Target} & R.A. \quad Declination & Parallax & ePlx & Spectral & Spectral & Var. \\
Name & HIP & Alias & (J2000)\textsuperscript{a} & \multicolumn{2}{c}{(mas)} & Type & Reference\textsuperscript{b} & Class\textsuperscript{c} \\
\hline
\endfirsthead
\multicolumn{9}{c}
{{\bfseries \tablename\ \thetable{} -- continued from previous page}} \\
\multicolumn{3}{c}{Target} & R.A. \quad Declination & Parallax & ePlx & Spectral & Spectral & Var. \\
Name & HIP & Alias & (J2000)\textsuperscript{a} & \multicolumn{2}{c}{(mas)} & Type & Reference\textsuperscript{b} & Class\textsuperscript{c} \\
\hline
\endhead

\hline \multicolumn{9}{r}{{Continued on next page}} \\
\endfoot

\hline
\caption*{\textsuperscript{a} Coordinates and parallaxes from~\citet{VanLeeuwen2007}, except for eta Car and VY CMa~\citep{Hog2000}, NML Tau~\citep{Skrutskie2006}, and RW LMi and TX Cam~\citep{Loup1993}.\\
\textsuperscript{b} References are meant to be representative: A26~\citep{A26}, B54~\citep{B54}, B85~\citep{B85}, C79~\citep{C79}, E57~\citep{E57}, E60~\citep{E60}, G48~\citep{G48}, G50~\citep{G50}, H58~\citep{H58}, H72~\citep{H72}, H75~\citep{H75}, J26~\citep{J26}, K42~\citep{K42}, K45~\citep{K45}, K74~\citep{K74}, K80~\citep{K80}, K89~\citep{K89}, L66~\citep{L66}, M43~\citep{M43}, M53~\citep{M53}, M73~\citep{M73}, P67~\citep{P67}, R52~\citep{R52}, S44~\citep{S44}, T08~\citep{T08}, W52~\citep{W52}, W57~\citep{W57}, W73~\citep{W73}, Y67~\citep{Y67}.\\
\textsuperscript{c} Variability Class is from the General Catalog of Variable Stars~\citep{Samus2009}.\\}
\endlastfoot

30 Psc & 154 & YY Psc & 00 01 57.62 -06 00 50.66 & 7.55 & 0.59 & M3 III & M73 & Lb: \\
bet And & 5447 & -- & 01 09 43.92 +35 37 14.01 & 16.52 & 0.56 & M0 III var & M43 & nsv \\
gam1 And & 9640 & -- & 02 03 53.95 +42 19 47.02 & 8.3 & 1.04 & K3 II & M73 & -- \\
omi Cet & 10826 & -- & 02 19 20.79 -02 58 39.50 & 10.91 & 1.22 & M5-M9e & J26 & Mira \\
alp Cet & 14135 & -- & 03 02 16.77 +04 05 23.06 & 13.09 & 0.44 & M1.5 IIIa & K89 & Lb: \\
rho Per & 14354 & -- & 03 05 10.59 +38 50 24.99 & 10.6 & 0.25 & M3 II-III & M73 & SRb \\
NML Tau & -- & IK Tau & 03 53 28.87 +11 24 21.70 & -- & -- & M9 & P67 & Mira \\
gam Eri & 18543 & -- & 03 58 01.77 -13 30 30.67 & 16.04 & 0.58 & M0 III-IIIb & K89 & Lb: \\
alp Tau & 21421 & -- & 04 35 55.24 +16 30 33.49 & 48.94 & 0.77 & K5 III & M43, R52 & Lb: \\
R Dor & 21479 & -- & 04 36 45.59 -62 04 37.80 & 18.31 & 0.99 & M8e & M73 & SRb \\
TX Cam & -- & -- & 05 00 50.39 +56 10 52.60 & -- & -- & M8-10 & W73 & Mira \\
RX Lep & 24169 & -- & 05 11 22.87 +05 09 02.75 & 6.71 & 0.44 & M6 III & A26 & SRb \\
alp Aur & 24608 & Capella & 05 16 41.35 +45 59 52.80 & 76.2 & 0.46 & G1 III+K0 III & Simbad & nsv \\
alp Ori & 27989 & -- & 05 55 10.31 +07 24 25.43 & 6.55 & 0.83 & M1.5 Iab & M73 & SRc \\
mu Gem & 30343 & -- & 06 22 57.63 +22 30 48.90 & 14.08 & 0.71 & M3 IIIab & K89 & Lb \\
alp Car & 30438 & -- & 06 23 57.11 -52 41 44.38 & 10.55 & 0.56 & F0 II & G48 & -- \\
alp CMa & 32349 & -- & 06 45 08.92 -16 42 58.02 & 379.21 & 1.58 & A1 V & M43 & nsv \\
L2 Pup & 34922 & -- & 07 13 32.32 -44 38 23.06 & 15.61 & 0.99 & M5 IIIe & B85 & SRb \\
VY CMa & 35793 & -- & 07 22 58.33 -25 46 03.24 & -- & -- & M5 Iae & H72 & -- \\
sig Pup & 36377 & -- & 07 29 13.83 -43 18 05.16 & 16.84 & 0.48 & K5 III & E57 & Ell: \\
alp CMi & 37279 & -- & 07 39 18.12 +05 13 29.96 & 284.56 & 1.26 & F5 IV & G48 & nsv \\
R Cnc & 40534 & -- & 08 16 33.83 +11 43 34.46 & 1.58 & 1.43 & M6.5-9e & K74 & Mira \\
lam Vel & 44816 & -- & 09 07 59.76 -43 25 57.33 & 5.99 & 0.11 & K4 Ib & K89 & Lc \\
RS Cnc & 45058 & -- & 09 10 38.80 +30 57 47.30 & 6.97 & 0.52 & M6S & B54 & SRc: \\
alp Hya & 46390 & -- & 09 27 35.24 -08 39 30.96 & 18.09 & 0.18 & K3 II-III & M73 & nsv \\
R Car & 46806 & -- & 09 32 14.60 -62 47 19.97 & 6.34 & 0.81 & M8e III: & K89 & Mira \\
R Leo & 48036 & -- & 09 47 33.49 +11 25 43.66 & 14.03 & 2.65 & M7-9e & K74 & Mira \\
CW Leo & -- & IRC+10216 & 09 47 57.41 +13 16 43.68 & -- & -- & C9 & Simbad & M \\
RW LMi & -- & CIT 6 & 10 16 02.27 +30 34 18.60 & -- & -- & C4,3e & C79 & SRa \\
U Ant & 51821 & -- & 10 35 12.85 -39 33 45.32 & 3.73 & 0.54 & C5,4 & S44 & Lb \\
eta Car & -- & -- & 10 45 03.55 -59 41 03.95 & -- & -- & O5.5 III - O7 I & T08 & S Dor \\
V Hya & 53085 & -- & 10 51 37.26 -21 15 00.32 & 1.44 & 1.41 & C7.5e & B54 & SRa \\
56 Leo & 53449 & VY Leo & 10 56 01.47 +06 11 07.33 & 8.39 & 0.37 & M5.5 III & M73 & Lb: \\
ome Vir & 56779 & -- & 11 38 27.61 +08 08 03.47 & 6.56 & 0.36 & M4.5: III & K89 & Lb \\
nu Vir & 57380 & -- & 11 45 51.56 +06 31 45.74 & 11.1 & 0.18 & M1 III & M73 & SRb \\
eps Mus & 59929 & -- & 12 17 34.28 -67 57 38.65 & 10.82 & 0.17 & M5 III & L66 & SRb: \\
gam Cru & 61084 & -- & 12 31 09.96 -57 06 47.57 & 36.83 & 0.18 & M3.5 III & K89 & nsv \\
del Vir & 63090 & -- & 12 55 36.21 +03 23 50.89 & 16.44 & 0.22 & M3 III & W57, H58 & nsv \\
R Hya & 65835 & -- & 13 29 42.78 -23 16 52.77 & 8.05 & 0.69 & M6.5-9e & K74 & Mira \\
W Hya & 67419 & -- & 13 49 02.00 -28 22 03.49 & 9.59 & 1.12 & M7.5-9e & K74 & SRa \\
2 Cen & 67457 & V806 Cen & 13 49 26.72 -34 27 02.79 & 17.82 & 0.21 & M4.5 IIIb & L66 & SRb \\
alp Boo & 69673 & -- & 14 15 39.67 +19 10 56.67 & 88.83 & 0.54 & K1.5 III & K89 & nsv \\
alp Cen B & 71681 & -- & 14 39 35.06 -60 50 15.10 & 796.92 & 25.9 & K1 V & H75 & -- \\
alp Cen A & 71683 & -- & 14 39 36.49 -60 50 02.37 & 754.81 & 4.11 & G2 V & G50 & -- \\
bet UMi & 72607 & -- & 14 50 42.33 +74 09 19.81 & 24.91 & 0.12 & K4 III var & R52, M53 & nsv \\
RR UMi & 73199 & -- & 14 57 35.01 +65 55 56.86 & 7.1 & 0.37 & M4.5 III & K89 & SRb \\
R Ser & 77615 & -- & 15 50 41.74 +15 08 01.10 & 4.78 & 1.84 & M5-7e & K74 & Mira \\
del Oph & 79593 & -- & 16 14 20.74 -03 41 39.56 & 19.06 & 0.16 & M0.5 III & M73 & nsv \\
U Her & 80488 & -- & 16 25 47.47 +18 53 32.86 & 4.26 & 0.85 & M8e & K74 & Mira \\
30 Her & 90704 & g Her & 16 28 38.55 +41 52 54.04 & 9.21 & 0.18 & M6 III & M73 & SRb \\
alp Sco & 80763 & -- & 16 29 24.46 -26 25 55.21 & 5.89 & 1.0 & M1.5 Iab & M73 & Lc \\
alp TrA & 82273 & -- & 16 48 39.89 -69 01 39.76 & 8.35 & 0.15 & K2 III & W52 & -- \\
alp Her & 84345 & -- & 17 14 38.86 +14 23 25.08 & 0.07 & 1.32 & M5 II & W57 & SRc \\
alp Her A & 84345 & -- & 17 14 38.86 +14 23 25.23 & 9.07 & 1.32 & M5 Ib-II & M73 & SRc \\
VX Sgr & 88838 & -- & 18 08 04.05 -22 13 26.63 & 3.82 & 2.73 & M4 Iae & H72 & SRc \\
eta Sgr & 89642 & -- & 18 17 37.64 -36 45 42.07 & 22.35 & 0.24 & M3 II & O60 & Lb: \\
alp Lyr & 91262 & -- & 18 36 56.33 +38 47 01.32 & 130.23 & 0.36 & A0 Va & Simbad & DSCTC \\
X Oph & 91389 & -- & 18 38 21.12 +08 50 02.75 & -- & -- & M0-8e + K2: III & K74 & Mira \\
R Lyr & 92862 & -- & 18 55 20.10 +43 56 45.93 & 10.94 & 0.12 & M5 III var & K45 & SRb \\
R Aql & 93820 & -- & 19 06 22.25 +08 13 48.01 & 2.37 & 0.87 & M6.5-9e & K74 & Mira \\
W Aql & -- & -- & 19 15 23.35 -07 02 50.35 & -- & -- & S6/6e & K80 & Mira \\
chi Cyg & 97629 & -- & 19 50 33.92 +32 54 50.61 & 5.53 & 1.1 & S6-9/1-2e & K80 & Mira \\
T Cep & 104451 & -- & 21 09 31.78 +68 29 27.20 & 5.33 & 0.9 & M6-9e & K74 & Mira \\
W Cyg & 106642 & -- & 21 36 02.50 +45 22 28.53 & 5.72 & 0.38 & M5+ IIIa & Y67 & SRb \\
mu Cep & 107259 & -- & 21 43 30.46 +58 46 48.16 & 0.55 & 0.2 & M2 Ia & K42 & SRc \\
pi1 Gru & 110478 & -- & 22 22 44.21 -45 56 52.61 & 6.13 & 0.76 & S5,7: & B54 & SRb \\
bet Gru & 112122 & -- & 22 42 40.05 -46 53 04.48 & 18.43 & 0.42 & M4.5 III & K89 & Lc: \\
lam Aqr & 112961 & -- & 22 52 36.86 -07 34 46.56 & 8.47 & 0.66 & M2 III & Simbad & LB \\
alp PsA & 113368 & -- & 22 57 39.05 -29 37 20.05 & 129.81 & 0.47 & A3 V & M43 & nsv \\
bet Peg & 113881 & -- & 23 03 46.46 +28 04 58.03 & 16.64 & 0.15 & M2.5 II-III & M73 & Lb \\
chi Aqr & 114939 & -- & 23 16 50.93 -07 43 35.40 & 5.32 & 0.37 & M3 III & Simbad & -- \\
R Aqr & 117054 & -- & 23 43 49.46 -15 17 04.14 & 2.76 & 2.27 & M6.5-9ep & K74 & Mira \\
R Cas & 118188 & -- & 23 58 24.87 +51 23 19.70 & 7.95 & 1.02 & M6-9e & K74 & Mira \\
\end{longtable}
\end{center}

%% file: caoss.tex
\begin{center}
\LTcapwidth=\textwidth
\begin{longtable}{ccccccc}
\caption{The Cassini Atlas of Stellar Spectra. All epochs listed are where flux calibrated spectra were recovered. The columns from left to right are: name of the stellar target, R.A. and Dec from Table~\ref{tab:starlist}, Cassini planning sequence, UTC and Julian dates, number of exposures, and cumulative exposure time per pixel. Sorted by Right Ascension. \label{tab:calibrated_data}}
\\ Target & R.A. \quad Declination & Sequence & \multicolumn{2}{c}{Epoch} & No. of & Total Exp.\\
 & (J2000) & & UTC & JD & Exp. & Time (s)\\
\hline
\endfirsthead
\multicolumn{7}{c}
{{\bfseries \tablename\ \thetable{} -- continued from previous page}} \\
Target & R.A. \quad Declination & Sequence & \multicolumn{2}{c}{Epoch} & No. of & Total Exp.\\
 & (J2000) & & UTC & JD & Exp. & Time (s)\\
\hline
\endhead

\hline \multicolumn{7}{r}{{Continued on next page}} \\
\endfoot

\hline
\endlastfoot

30 Psc & 00 01 57.62 -06 00 50.66 & S25 & 2006-11-5 & 2454044 & 15 & 4.7 \\
bet And & 01 09 43.92 +35 37 14.01 & S18 & 2006-2-1 & 2453767 & 15 & 4.7 \\
gam1 And & 02 03 53.95 +42 19 47.02 & S18 & 2006-2-1 & 2453767 & 15 & 4.7 \\
omi Cet & 02 19 20.79 -02 58 39.50 & S83 & 2014-4-11 & 2456758 & 29 & 18.1 \\
alp Cet & 03 02 16.77 +04 05 23.06 & S18 & 2006-2-1 & 2453767 & 14 & 4.1 \\
rho Per & 03 05 10.59 +38 50 24.99 & S18 & 2006-2-1 & 2453767 & 21 & 5.0 \\
NML Tau & 03 53 28.87 +11 24 21.70 & C33 & 2002-7-19 & 2452474 & 22 & 15.1 \\
gam Eri & 03 58 01.77 -13 30 30.67 & S27 & 2007-2-16 & 2454147 & 14 & 4.1 \\
alp Tau & 04 35 55.24 +16 30 33.49 & C35 & 2003-1-19 & 2452658 & 40 & 20.2 \\
R Dor & 04 36 45.59 -62 04 37.80 & C33 & 2002-7-18 & 2452473 & 82 & 54.3 \\
TX Cam & 05 00 50.39 +56 10 52.60 & S27 & 2007-2-15 & 2454146 & 16 & 5.4 \\
RX Lep & 05 11 22.87 +05 09 02.75 & S27 & 2007-2-16 & 2454147 & 14 & 4.1 \\
alp Ori & 05 55 10.31 +07 24 25.43 & S18 & 2006-2-1 & 2453767 & 20 & 1.9 \\
mu Gem & 06 22 57.63 +22 30 48.90 & S18 & 2006-2-1 & 2453767 & 19 & 3.1 \\
alp Car & 06 23 57.11 -52 41 44.38 & C34 & 2002-10-19 & 2452566 & 39 & 7.8 \\
alp Car & 06 23 57.11 -52 41 44.38 & S69 & 2011-7-27 & 2455769 & 260 & 44.7 \\
alp CMa & 06 45 08.92 -16 42 58.02 & C39 & 2003-10-13 & 2452925 & 43 & 12.9 \\
alp CMa & 06 45 08.92 -16 42 58.02 & S69 & 2011-7-27 & 2455769 & 15 & 8.3 \\
L2 Pup & 07 13 32.32 -44 38 23.06 & S36 & 2008-1-12 & 2454477 & 14 & 2.2 \\
VY CMa & 07 22 58.33 -25 46 03.24 & S36 & 2008-1-12 & 2454477 & 7 & 2.5 \\
sig Pup & 07 29 13.83 -43 18 05.16 & S83 & 2014-5-1 & 2456778 & 52 & 30.7 \\
alp CMi & 07 39 18.12 +05 13 29.96 & S08 & 2005-1-30 & 2453400 & 15 & 4.7 \\
R Cnc & 08 16 33.83 +11 43 34.46 & S79 & 2013-7-5 & 2456478 & 48 & 30.5 \\
lam Vel & 09 07 59.76 -43 25 57.33 & S36 & 2008-1-12 & 2454477 & 9 & 2.6 \\
RS Cnc & 09 10 38.80 +30 57 47.30 & S36 & 2008-1-12 & 2454477 & 14 & 2.8 \\
RS Cnc & 09 10 38.80 +30 57 47.30 & S79 & 2013-7-5 & 2456478 & 48 & 30.5 \\
alp Hya & 09 27 35.24 -08 39 30.96 & S35 & 2007-12-7 & 2454441 & 11 & 4.1 \\
alp Hya & 09 27 35.24 -08 39 30.96 & S38 & 2008-3-9 & 2454534 & 8 & 3.2 \\
alp Hya & 09 27 35.24 -08 39 30.96 & S81 & 2013-12-20 & 2456646 & 160 & 50.6 \\
R Car & 09 32 14.60 -62 47 19.97 & S38 & 2008-2-27 & 2454523 & 6 & 1.9 \\
R Car & 09 32 14.60 -62 47 19.97 & S40 & 2008-4-26 & 2454582 & 26 & 10.1 \\
R Car & 09 32 14.60 -62 47 19.97 & S52 & 2009-8-1 & 2455044 & 32 & 12.7 \\
R Leo & 09 47 33.49 +11 25 43.66 & C37 & 2003-5-19 & 2452778 & 47 & 9.5 \\
R Leo & 09 47 33.49 +11 25 43.66 & S38 & 2008-3-3 & 2454528 & 20 & 5.4 \\
R Leo & 09 47 33.49 +11 25 43.66 & S40 & 2008-4-26 & 2454582 & 30 & 10.5 \\
R Leo & 09 47 33.49 +11 25 43.66 & S81 & 2013-12-20 & 2456646 & 160 & 50.6 \\
RW LMi & 10 16 02.27 +30 34 18.60 & S08 & 2005-1-30 & 2453400 & 9 & 3.3 \\
U Ant & 10 35 12.85 -39 33 45.32 & S84 & 2014-6-29 & 2456837 & 117 & 74.4 \\
eta Car & 10 45 03.55 -59 41 03.95 & C33 & 2002-7-18 & 2452473 & 58 & 39.0 \\
eta Car & 10 45 03.55 -59 41 03.95 & S38 & 2008-2-27 & 2454523 & 8 & 3.2 \\
eta Car & 10 45 03.55 -59 41 03.95 & S81 & 2013-12-27 & 2456653 & 85 & 54.1 \\
V Hya & 10 51 37.26 -21 15 00.32 & S28 & 2007-2-25 & 2454156 & 35 & 10.6 \\
56 Leo & 10 56 01.47 +06 11 07.33 & S35 & 2007-12-7 & 2454441 & 14 & 4.1 \\
ome Vir & 11 38 27.61 +08 08 03.47 & S28 & 2007-3-14 & 2454173 & 21 & 7.0 \\
ome Vir & 11 38 27.61 +08 08 03.47 & S38 & 2008-3-9 & 2454534 & 29 & 16.5 \\
nu Vir & 11 45 51.56 +06 31 45.74 & S28 & 2007-3-14 & 2454173 & 21 & 8.4 \\
eps Mus & 12 17 34.28 -67 57 38.65 & S28 & 2007-2-25 & 2454156 & 26 & 7.2 \\
eps Mus & 12 17 34.28 -67 57 38.65 & S38 & 2008-2-27 & 2454523 & 9 & 3.2 \\
eps Mus & 12 17 34.28 -67 57 38.65 & S52 & 2009-8-1 & 2455044 & 36 & 12.8 \\
gam Cru & 12 31 09.96 -57 06 47.57 & C35 & 2003-1-19 & 2452658 & 10 & 6.4 \\
gam Cru & 12 31 09.96 -57 06 47.57 & S38 & 2008-2-28 & 2454524 & 28 & 4.4 \\
gam Cru & 12 31 09.96 -57 06 47.57 & S52 & 2009-8-1 & 2455044 & 48 & 13.2 \\
del Vir & 12 55 36.21 +03 23 50.89 & S08 & 2005-1-30 & 2453400 & 16 & 4.2 \\
R Hya & 13 29 42.78 -23 16 52.77 & C37 & 2003-5-18 & 2452777 & 5 & 2.2 \\
R Hya & 13 29 42.78 -23 16 52.77 & S38 & 2008-3-19 & 2454544 & 14 & 2.2 \\
W Hya & 13 49 02.00 -28 22 03.49 & C40 & 2003-12-1 & 2452974 & 59 & 17.9 \\
W Hya & 13 49 02.00 -28 22 03.49 & S38 & 2008-3-19 & 2454544 & 16 & 3.5 \\
2 Cen & 13 49 26.72 -34 27 02.79 & S35 & 2007-11-14 & 2454418 & 18 & 1.8 \\
alp Boo & 14 15 39.67 +19 10 56.67 & C25 & 2001-3-26 & 2451994 & 44 & 6.9 \\
alp Boo & 14 15 39.67 +19 10 56.67 & C27 & 2001-8-29 & 2452150 & 258 & 40.2 \\
alp Boo & 14 15 39.67 +19 10 56.67 & C34 & 2002-10-13 & 2452560 & 2 & 0.4 \\
alp Boo & 14 15 39.67 +19 10 56.67 & S16 & 2005-12-2 & 2453706 & 7 & 1.1 \\
alp Boo & 14 15 39.67 +19 10 56.67 & S36 & 2008-1-5 & 2454470 & 73 & 2.6 \\
alp Boo & 14 15 39.67 +19 10 56.67 & S53 & 2009-8-29 & 2455072 & 25 & 2.1 \\
alp Cen & 14 39 36.49 -60 50 02.37 & C40 & 2003-12-1 & 2452974 & 50 & 12.2 \\
alp Cen & 14 39 36.49 -60 50 02.37 & S38 & 2008-3-19 & 2454544 & 15 & 3.4 \\
alp Cen & 14 39 36.49 -60 50 02.37 & S51 & 2009-7-2 & 2455014 & 338 & 215.0 \\
alp Cen & 14 39 36.49 -60 50 02.37 & S52 & 2009-8-1 & 2455044 & 39 & 12.9 \\
alp Cen & 14 39 36.49 -60 50 02.37 & S72 & 2012-3-29 & 2456015 & 212 & 67.0 \\
bet UMi & 14 50 42.33 +74 09 19.81 & S82 & 2014-2-20 & 2456708 & 45 & 28.6 \\
RR UMi & 14 57 35.01 +65 55 56.86 & S82 & 2014-2-20 & 2456708 & 45 & 28.6 \\
R Ser & 15 50 41.74 +15 08 01.10 & S78 & 2013-4-15 & 2456397 & 121 & 77.0 \\
del Oph & 16 14 20.74 -03 41 39.56 & S26 & 2007-1-4 & 2454104 & 15 & 4.7 \\
del Oph & 16 14 20.74 -03 41 39.56 & S53 & 2009-8-29 & 2455072 & 14 & 3.8 \\
U Her & 16 25 47.47 +18 53 32.86 & S78 & 2013-4-15 & 2456397 & 111 & 70.6 \\
30 Her & 16 28 38.55 +41 52 54.04 & S22 & 2006-7-29 & 2453945 & 33 & 4.2 \\
30 Her & 16 28 38.55 +41 52 54.04 & S53 & 2009-8-29 & 2455072 & 24 & 2.7 \\
alp Sco & 16 29 24.46 -26 25 55.21 & C33 & 2002-7-12 & 2452467 & 73 & 48.2 \\
alp Sco & 16 29 24.46 -26 25 55.21 & S31 & 2007-7-8 & 2454289 & 470 & 16.9 \\
alp Sco & 16 29 24.46 -26 25 55.21 & S51 & 2009-7-17 & 2455029 & 68 & 26.4 \\
alp TrA & 16 48 39.89 -69 01 39.76 & S08 & 2005-1-30 & 2453400 & 12 & 2.8 \\
alp TrA & 16 48 39.89 -69 01 39.76 & S52 & 2009-8-1 & 2455044 & 32 & 12.7 \\
alp Her & 17 14 38.86 +14 23 25.08 & S13 & 2005-8-28 & 2453610 & 17 & 2.4 \\
alp Her & 17 14 38.86 +14 23 25.08 & S53 & 2009-8-29 & 2455072 & 28 & 2.9 \\
VX Sgr & 18 08 04.05 -22 13 26.63 & S51 & 2009-7-17 & 2455029 & 34 & 14.4 \\
eta Sgr & 18 17 37.64 -36 45 42.07 & S22 & 2006-7-30 & 2453946 & 16 & 4.2 \\
eta Sgr & 18 17 37.64 -36 45 42.07 & S51 & 2009-7-17 & 2455029 & 38 & 15.9 \\
X Oph & 18 38 21.12 +08 50 02.75 & S26 & 2007-1-4 & 2454104 & 26 & 5.1 \\
X Oph & 18 38 21.12 +08 50 02.75 & S60 & 2010-5-26 & 2455342 & 7 & 2.2 \\
R Lyr & 18 55 20.10 +43 56 45.93 & S13 & 2005-8-28 & 2453610 & 24 & 4.5 \\
R Lyr & 18 55 20.10 +43 56 45.93 & S79 & 2013-6-20 & 2456463 & 11 & 1.7 \\
R Aql & 19 06 22.25 +08 13 48.01 & S60 & 2010-5-26 & 2455342 & 6 & 1.9 \\
R Aql & 19 06 22.25 +08 13 48.01 & S69 & 2011-7-27 & 2455769 & 225 & 42.3 \\
W Aql & 19 15 23.35 -07 02 50.35 & S26 & 2007-1-4 & 2454104 & 20 & 4.9 \\
chi Cyg & 19 50 33.92 +32 54 50.61 & S22 & 2006-7-29 & 2453945 & 33 & 5.3 \\
T Cep & 21 09 31.78 +68 29 27.20 & S13 & 2005-8-28 & 2453610 & 22 & 4.4 \\
T Cep & 21 09 31.78 +68 29 27.20 & S78 & 2013-5-29 & 2456441 & 92 & 58.5 \\
W Cyg & 21 36 02.50 +45 22 28.53 & S79 & 2013-6-21 & 2456464 & 98 & 46.3 \\
mu Cep & 21 43 30.46 +58 46 48.16 & S22 & 2006-7-30 & 2453946 & 35 & 5.5 \\
pi1 Gru & 22 22 44.21 -45 56 52.61 & S22 & 2006-7-30 & 2453946 & 28 & 4.4 \\
bet Gru & 22 42 40.05 -46 53 04.48 & C34 & 2002-10-5 & 2452552 & 40 & 7.8 \\
alp PsA & 22 57 39.05 -29 37 20.05 & C25 & 2001-3-28 & 2451996 & 12 & 7.6 \\
alp PsA & 22 57 39.05 -29 37 20.05 & C34 & 2002-10-19 & 2452566 & 32 & 20.4 \\
alp PsA & 22 57 39.05 -29 37 20.05 & C35 & 2003-1-22 & 2452661 & 13 & 8.3 \\
alp PsA & 22 57 39.05 -29 37 20.05 & S72 & 2012-3-30 & 2456016 & 10 & 6.4 \\
bet Peg & 23 03 46.46 +28 04 58.03 & S13 & 2005-8-28 & 2453610 & 24 & 4.5 \\
R Aqr & 23 43 49.46 -15 17 04.14 & S22 & 2006-7-30 & 2453946 & 15 & 4.7 \\
R Cas & 23 58 24.87 +51 23 19.70 & S08 & 2005-1-30 & 2453400 & 21 & 5.6 \\
\end{longtable}
\end{center}